\documentstyle[twocolumn,seceq]{jpsj}

\newcommand{\be}{\begin{equation}}
\newcommand{\ee}{\end{equation}}
\newcommand{\bea}{\begin{eqnarray}}
\newcommand{\eea}{\end{eqnarray}}
\newcommand{\lsim}{\raise.35ex\hbox{$<$}\kern-0.75em\lower.5ex\hbox{$\sim$}}
\newcommand{\gsim}{\raise.35ex\hbox{$>$}\kern-0.75em\lower.5ex\hbox{$\sim$}}
\def\dfrac#1#2{{\displaystyle\frac{#1}{#2}}}

\def\runtitle{Dynamical Properties of the $1/r^2$-Type Supersymmetric 
{\it t-J} Model in a Magnetic Field}
\def\runauthor{Yasuhiro {\sc Saiga} and Yoshio {\sc Kuramoto}}

\title
{
Dynamical Properties of the 1/\mbox{\boldmath $r^2$}-Type Supersymmetric
\mbox{\boldmath $t$}-\mbox{\boldmath $J$} Model in a Magnetic Field:  
Manifestation of Spin-Charge Separation
}

\author
{ 
Yasuhiro {\sc Saiga}\footnote{E-mail: saiga@issp.u-tokyo.ac.jp}
and Yoshio {\sc Kuramoto}$^1$
}

\inst
{
Institute for Solid State Physics, University of Tokyo,\\
Kashiwa-no-ha 5-1-5, Kashiwa 277-8581 \\
$^1$Department of Physics, Tohoku University, Sendai 980-8578
}

\recdate
{\hspace{2cm}}

\abst
{
Quasi-particle picture in a magnetic field
is pursued for dynamical spin and charge
correlation functions of the one-dimensional supersymmetric {\it t-J}
model with inverse-square interaction. 
With use of exact diagonalization and the asymptotic Bethe-ansatz
equations for finite systems, excitation contents of relevant excited
states are identified 
which are valid in the thermodynamic limit.
The excitation contents are composed of spinons, antispinons, holons
and antiholons obeying fractional statistics. 
Both longitudinal and transverse components of the dynamical spin
structure factor are independent of the electron density in the region
where only quasi-particles with spin degrees of freedom (spinons and
antispinons) contribute.
The dynamical charge structure factor does not depend on the
spin-polarization density in the region where only quasi-particles
with charge (holons and antiholons) are excited.
These features indicate the strong spin-charge separation in 
dynamics, reflecting the high symmetry of the model.
}

\kword
{
dynamical spin structure factor, dynamical charge structure factor,
supersymmetric {\it t-J} model, inverse-square interaction,
magnetic field, spin-charge separation, exact diagonalization, 
recursion method
}

\begin{document}
\sloppy
\maketitle

%\newpage

%%%%%%%%%%%%%%%%%%%%%%%%%%%%%%%%%%%%%%%%%%%%%%%%%%%%%%%%%%%%%%%%%%%%%%%%%%%

\section{Introduction}

The concept of spin-charge separation 
is no longer a purely academic idea.  
In various low-dimensional electron systems,
angle-resolved photoemission measurements have revealed the
existence of spin-charge separation.
In quasi-one-dimensional (1D) Mott insulators such as  
SrCuO$_2$~\cite{Kim96,Kim97} and
Sr$_2$CuO$_3$,~\cite{Fujisawa98,Fujisawa99} 
angle-resolved photoemission data show
two distinct dispersions in the energy-momentum plane.
The two correspond to the spinon band and the holon band, respectively.
This is an indication of the spin-charge separation.
It has been reported that the spin-charge separation in Sr$_2$CuO$_3$ 
appears also 
in dielectric response.\cite{Neudert}
\par

Recently magnetic-field-induced dimensional crossover has been
observed in the normal state of YBa$_2$Cu$_4$O$_8$.~\cite{Hussey}
The CuO chains of this material play the critical role in magnetoconductivity. 
The following question may arise:
When a 1D electron system is under a magnetic field, 
how is the spin-charge separation affected by the field?
The magnetic field controls spin degrees of freedom, while
the hole doping controls charge degrees of freedom.
However, realistic situation should be rather complex due to
convolution of both degrees of freedom in a strongly correlated system.
\par

It is well-known that interacting electron systems in one dimension are 
described by the Tomonaga-Luttinger liquid,~\cite{Solyom,Haldane81}
but the information is restricted to that in the long-wavelength 
and low-energy limit.
One can adopt the Hubbard model and the supersymmetric {\it t-J} model
as Bethe-ansatz solvable models.
With use of the exact wave function of the $U \to \infty$ Hubbard
model, critical exponents and global features of static spin structure
factors have been investigated in a magnetic field.~\cite{FK1,FK2,OSS}
The exponents depend only on the magnetization per electron and 
are independent of the electron density.
For the supersymmetric {\it t-J} model with nearest-neighbor
interaction, the exponents depend both on the magnetization and the
electron density.~\cite{KawakamiYang}
In order to obtain deeper understanding of spin-charge coupling, 
it is imperative to derive theoretically not only
the critical exponents but also the overall spectral weight in dynamics.
\par

The supersymmetric {\it t-J} model with inverse-square
interaction~\cite{KY} reveals the spin-charge separation in a typical
form.
The spin velocity is independent of the electron density $\bar{n}$,
and the charge velocity is independent of the spin-polarization
density $\bar{m}$.~\cite{Kawakami}
At low temperature, the spin susceptibility is independent of
$\bar{n}$, and the charge susceptibility is independent of
$\bar{m}$.~\cite{KK1,KK2}
It has been found from the exact thermodynamics that 
elementary excitations consist of four types of
quasi-particles: spinons, antispinons, holons and antiholons.~\cite{KK1}
In zero magnetic field, the regions of nonvanishing spectral weight
in the momentum-energy plane, which are called the ``compact 
support'', have been obtained for various dynamical correlation 
functions at zero temperature.~\cite{HH}
The excitation contents are composed of spinons, holons and
antiholons.
In particular, the weight of the dynamical spin structure factor 
does not depend on $\bar{n}$ in the region
where only two spinons contribute.~\cite{SK}
At half-filling 
with $\bar{n}=1$, i.e., 
in the Haldane-Shastry model,~\cite{Haldane88,Shastry88} 
magnetic-field dynamics has been investigated.~\cite{TH}
So far, study of full dynamics are lacking at arbitrary filling and
spin-polarization density.
\par

In this paper, we extend our previous work~\cite{SK} to the case of finite
magnetic field for the $1/r^2$-type supersymmetric {\it t-J} model. 
We investigate how the spin-charge separation appears
in the spin and charge dynamics in {\it nonzero} field.
To this end, exact diagonalization up to 16 sites is used.
In order to analyze the results of dynamical quantities, 
we employ the asymptotic Bethe-ansatz equations~\cite{Kawakami,WLC}
and the skew Young diagram.~\cite{SK,KKN,HB}
In addition, we examine the static structure factors 
in the $1/r^2$ model as well as the {\it t-J} model 
with realistic parameters.
\par

This paper is organized as follows:
Section 2 describes the model and
definitions of dynamical quantities.
In $\S$3 we derive the supports for dynamical spin and charge 
structure factors.
In $\S$4 we discuss the spectral weights themselves,
and clarify manifestation of the strong spin-charge separation
in dynamics.
In $\S$5 we show the results of the static structure factors. 
Finally we give the summary of this study in $\S$6. 
A brief account of this work has been presented in ref.\ 25.
\par

%%%%%%%%%%%%%%%%%%%%%%%%%%%%%%%%%%%%%%%%%%%%%%%%%%%%%%%%%%%%%%%%%%%%%%%%%%%

\section{Model and Dynamical Structure Factors}

We consider the following Hamiltonian:
\bea
  {\cal H} &=& \sum_{i < j}^N \Biggl[ - \sum_{\sigma} t I_{ij}
\left( \tilde{c}_{i \sigma}^{\dagger} \tilde{c}_{j \sigma}
+ \tilde{c}_{j \sigma}^{\dagger} \tilde{c}_{i \sigma}
    \right) \nonumber \\
           & & + J I_{ij} \left( \mbox{\boldmath $S$}_i \cdot
\mbox{\boldmath $S$}_j - \frac{1}{4} n_i n_j \right) \Biggr] 
- g \mu_{\rm B} H \sum_{i=1}^N S_i^z,\qquad
\eea
with even number $N$ of sites and average electron number $\bar{n}$ 
per site.
Here $\tilde{c}_{i \sigma} = c_{i \sigma} (1 - n_{i,- \sigma})$ is the 
annihilation operator of an electron with
spin $\sigma$ at site $i$ with the constraint of no double occupation.
The Zeeman term couples the system to a uniform magnetic field $H$ 
in the $z$ direction.
A dimensionless parameter $h$ is defined as $h = g \mu_{\rm B} H/t$.
We assume the periodic boundary condition.
Then $I_{ij}$ is given by
$I_{ij} = [(N/\pi) \sin(\pi (i - j)/N)]^{-2}$ for the $1/r^2$ type, 
and by $I_{ij} = \delta_{j,i+1}$ for the nearest-neighbor type.
In this paper we mainly discuss the $1/r^2$-type supersymmetric 
{\it t-J} model ($J/t = 2$).
\par

The dynamical spin and charge structure factors are given by
\bea
  S^{\alpha \alpha}(q,\omega) &=& \sum_\nu 
\left| \langle \nu |s_q^{\alpha}| 0 \rangle
\right|^2 \delta(\omega -E_\nu +E_0), \\
  N(q,\omega) &=& \sum_\nu \left| \langle \nu |n_q| 0 \rangle
\right|^2 \delta(\omega -E_\nu +E_0),
\eea
where $s_q^{\alpha} = N^{-1/2} \sum_\ell s_\ell^{\alpha} {\rm e}^{- {\rm i}
q \ell}$ ($\alpha = x, y, z$), 
$n_q = N^{-1/2} \sum_\ell (n_\ell - \bar{n}) {\rm
e}^{- {\rm i} q \ell}$, and $|\nu \rangle$ denotes an eigenstate of
${\cal H}$ with energy $E_{\nu}$ ($E_0$ being the ground-state energy).
The transverse component $S^{xx}(q,\omega) \equiv S^{yy}(q,\omega)$ is
decomposed into two parts: 
\be
  S^{xx}(q,\omega) = S^{yy}(q,\omega) 
= \frac{1}{4} \left[ S^{-+}(q,\omega) + S^{+-}(q,\omega) \right].
\ee
We calculate $S^{-+}(q,\omega)$ and $S^{+-}(q,\omega)$ instead of 
$S^{xx}(q,\omega)$ and $S^{yy}(q,\omega)$.
These dynamical quantities can be written in the form of a
continued fraction.~\cite{GB}
The coefficients of the continued fraction are obtained from 
the Lanczos algorithm.
This method is called the recursion method.
We truncate the continued fraction 
between 100 and 200 iterations 
and take the Lorentzian width 
of ${\cal O}(10^{-5}t)$.
In our calculation, energy levels and intensities have accuracies of 
7 digits and 4-5 digits, respectively.
\par

We assume $\bar{m} \geq 0$ without loss of generality.
In the thermodynamic limit,  $\bar{m}$ 
is determined by $h$ as follows:
\be
  \bar{m} = \left\{
   \begin{array}{ll}
    1 - \sqrt{1 - 2 h/\pi^2}, &\mbox{for $0 \le h \le h_{\rm c}$},\\
    \bar{n} &\mbox{for $h \ge h_{\rm c}$}
   \end{array}\right.
\ee
where $h_{\rm c} = \bar{n} (2 - \bar{n}) \pi^2/2$.~\cite{Kawakami}.
For finite systems, $\bar{m}=(N-Q-2M)/N$ and $\bar{n}=(N-Q)/N$.
Here $Q$ and $M$ 
denote the numbers of holes and down-spins, respectively.
For $\bar{n} < 1$ we consider the case of $Q=$ even and $M=$ odd  
under the periodic boundary condition.
\par

%%%%%%%%%%%%%%%%%%%%%%%%%%%%%%%%%%%%%%%%%%%%%%%%%%%%%%%%%%%%%%%%%%%%%%%%%%%

\section{Spectra of Quasi-Particles and Supports}

In this section we determine the regions of nonvanishing spectral weight 
in the energy-momentum plane (i.e., the compact supports) 
for various dynamical correlation functions.
To achieve 
this goal, the following points should be clarified:
\begin{itemize}
\item Dispersion relations of the relevant quasi-particles.
\item Excitation contents of the excited states.
\item Selection rules for the excited states.
\end{itemize}
Let us begin with the dispersion relations.
\par

\subsection{Dispersion relations of quasi-particles}

The spectra of quasi-particles in the model are most easily obtained 
by the asymptotic Bethe ansatz method.  
It is known~\cite{Kawakami,WLC,Haldane94}
that the method gives a part of the exact eigenstates corresponding to 
the Yangian highest-weight states (YHWS), 
but that other eigenstates are missed.
Since the YHWS contain all kinds of quasi-particles,
the asymptotic Bethe ansatz is sufficient to derive the dispersion relations.
The Yangian multiplet structure of eigenstates is then derived by 
the skew Young diagram.\cite{SK} 

We write the energy of a spinon with spin $\sigma$  as 
$\epsilon_{\rm s}^{\sigma}$, while 
that of an antispinon as $\epsilon_{\bar{\rm s}}^{\downarrow}$.  
Note that an antispinon can have only a down spin with $\bar{m}\geq 0$.
The energies of a holon and an antiholon are written as 
$\tilde{\epsilon}_{\rm h}$ and $\tilde{\epsilon} _{\bar{\rm h}}$, 
respectively.  
Details of the derivation of the spectra are given in  Appendix.
The final results are given by
\bea
  & & \epsilon_{\rm s}^{\uparrow}/t = - k (k - \pi) - \frac{h}{2},
\label{upspinondispersion} \\
  & & \epsilon_{\rm s}^{\downarrow}/t = - k (k - \pi) + \frac{h}{2},
\label{downspinondispersion} \\
  & & \epsilon_{\bar{\rm s}}^{\downarrow}/t = 
\frac{1}{2} k (k - 2 \pi) + h,\label{downantispinondispersion} \\
  & & 
\tilde{\epsilon}_{\rm h}/t = \left( k - \frac{\pi}{2} \right)^2 + 
\frac{\pi^2}{12},\label{holondispersion2} \\
  & & 
\tilde{\epsilon} _{\bar{\rm h}}/t = 
- \frac{1}{2} \left( k - \pi \right)^2 - \frac{\pi^2}{6},
\label{antiholondispersion2}
\eea
where $h = \bar{m} (2 - \bar{m}) \pi^2/2$.

We note that only a part of the momentum space is relevant 
as shown in Fig.\ \ref{fig.spectra}.
For classification of excited states, it is convenient to define the right and left branches for 
each kind of quasi-particles.
For example, the right [left] up and down spinons are allowed only in 
the range
$\bar{m} \pi/2 \le k \le \bar{n} \pi/2$ 
[$(1 - \bar{n}/2) \pi \le k \le (1 - \bar{m}/2) \pi$]; 
the right [left] holons are allowed only 
for 
$(1 - \bar{n}/2) \pi \le k \le (1 - \bar{m}/2) \pi$ 
[$\bar{m} \pi/2 \le k \le \bar{n} \pi/2$].
The range of momentum for the right [left] antispinons is 
$(2 - \bar{m}) \pi \le k \le 2 \pi$ [$0 \le k \le \bar{m} \pi$], 
while the antiholons propagate in the region 
$\bar{n} \pi \le k \le (2 - \bar{n}) \pi$.

By shifting the origins of momentum and energy 
for each kind of quasi-particles, 
we can write the dispersion relations for right (left) spinons $\epsilon_{{\rm s}_{\rm R(L)}}$, 
right (left) antispinons $\epsilon_{\bar{\rm s}_{\rm R(L)}}$, right (left) holons $\epsilon_{{\rm h}
_{\rm R(L)}}$
and antiholons $\epsilon_{\bar{\rm h}}$ in the following form:
\bea
  & & \epsilon_{{\rm s}_{\rm R(L)}}/t = - k (k \mp v_{\rm s}^0),
\label{spinondispersion} \\
  & & \epsilon_{\bar{\rm s}_{\rm R(L)}}/t = 
\frac{1}{2} k (k \pm 2 v_{\rm s}^0),\label{antispinondispersion} \\
  & & \epsilon_{{\rm h}_{\rm R(L)}}/t = k (k \pm v_{\rm c}^0),
\label{holondispersion} \\
  & & \epsilon_{\bar{\rm h}}/t = 
\frac{1}{2} \left[ \left( v_{\rm c}^0 \right)^2 - k^2 \right],
\label{antiholondispersion}
\eea
where $v_{\rm s}^0 = (1 - \bar{m}) \pi$ and $v_{\rm c}^0 = (1 - \bar{n}) \pi$.
In the shifted momentum space, 
the right [left] spinons and holons are allowed in 
the range
$0 \le k \le (\bar{n} - \bar{m}) \pi/2$ 
[$- (\bar{n} - \bar{m}) \pi/2 \le k \le 0$].
The 
allowed region of momentum for the right (left) antispinons is 
$0 \le k \le \bar{m} \pi$ ($- \bar{m} \pi \le k \le 0$), 
while the antiholons propagate in the region 
$- v_{\rm c}^0 \le k \le v_{\rm c}^0$.
In the case of $\bar{m} = 0$,
these dispersion relations and the range of momentum reduce 
to those given by Ha and Haldane~\cite{HH}.
\par

\subsection{Determination of the supports} \label{Dsupport}

The remaining task to determine the supports is to derive 
excitation contents and selection rules for each dynamical 
correlation function.
\par

First, let us consider excitation contents.
For a given magnetic field, the magnetization of the ground state is 
fixed.
In the subspace of relevant excited states, 
we use the solution of total energy and total momentum 
from the asymptotic Bethe-ansatz equations.~\cite{Kawakami,WLC}
We then obtain the motif corresponding to the ground state and
excited states with finite intensity obtained by the recursion method.
The motif of the ground state with finite magnetization is 
given by
$\underbrace{00...00}_{\rm I}|\underbrace{1010...10}_{\rm II}|
\underbrace{11...11}_{\rm III}|\underbrace{01...0101}_{\rm IV}|
\underbrace{00...00}_{\rm V}$.
This ground-state motif turns out to be divided into five segments (I-V).
The segment III can be regarded as a holon condensate,
and the segments I and V can be regarded as a up-spinon condensate.
As a perturbation from the magnetized ground state, 
spinons (00) and holons (11) are excited in the segments II and/or IV;
Antiholons (0) are created in the segment III, 
and antispinons (1) are created in the segments I and/or V.
Spinons and holons have semionic statistics, 
while antispinons and antiholons have Bose statistics.

Let $S^z$ and $C$ denote the $z$-components of spin 
and the charge of each quasi-particle, respectively.
It is natural to assign $(S^z, C) = (1/2, 0)$ to the up spinon, 
$(-1/2, 0)$ to the down spinon, 
$(-1, 0)$ to the down antispinon, 
$(0, +{\rm e})$ to the holon and 
$(0, -2{\rm e})$ to the antiholon~\cite{HH,TH,comment}.
However, this assignment leads to cases 
where some states with excitation contents decided above 
do not fulfill the spin and/or charge conservations 
for the excited states.
Moreover, those states do not fulfill the statistics of quasi-particles, 
either.
This situation has occurred also in zero magnetic field.\cite{SK}
\par

In ref.\ 18, we introduced the skew Young diagram in the $1/r^2$-type
supersymmetric {\it t-J} model.
By using this diagram, one can reproduce the supermultiplet structure,
i.e., 
nontrivial degeneracies in the excitation spectrum.
In addition, excitation contents for a finite system can be extracted
by the skew Young diagram.
This idea can be expanded into the magnetized case.
We first determine the five segments of a diagram from 
the corresponding motif.
Note that the segment III consists of $Q$ boxes, 
and that each of the segments I and V 
consists of $(N-Q-2M)/2$ boxes.
Next, we put an index for the internal degrees of freedom on 
each box of a skew Young diagram, according to the rules presented 
in ref.\ 18.
We represent an up-spin by 1, a down-spin by 2, and a hole by $\circ$.
One can read quasi-particles from a diagram of an excited state
as follows:

(i) We regard $\circ$ in the segments II and IV as a holon.

(ii) We regard a row of a 1-2 pair in the segment III as an antiholon.

(iii) In the segments II and IV, single 1 or 2 in a column, both of which 
do not make a pair, is regarded as a spinon; 
it does not matter whether the column include $\circ$ or not.
Single 1 corresponds to an up-spinon, and single 2 to a down spinon.

(iv) A column of a 1-2 pair in the segments I and V is regarded 
as an antispinon.

Because an antiholon or an antispinon consists of two boxes,
there exist cases where only a part of a 1-2 pair is
present in the segment defined by (ii) and (iv).
We shall call such a part of the 1-2 pair a ``half-antiholon 
($\bar{\rm h}^\ast$)'' 
for the former, and a ``half-antispinon
($\bar{\rm s}_{{\rm L}({\rm R})}^\ast$)'' 
for the latter. 
The half-antiholon [half-antispinon] is assumed to have 
$(S^z, C) = (0, -{\rm e})$ [$(-1/2, 0)$] and semionic statistics.
Thereby the excitation contents from the diagram satisfies both
charge and spin conservations and the statistical feature.

Let us give a example of the skew Young diagram.
For $S^{+-}(q,\omega)$ from the initial state with $(N,Q,M)=(12,2,3)$, 
the excitation energy with $(q/\pi,\omega/t)=(0.5,2.05617)$ has 
finite intensity.
Its motif is $010|10|110|10|100$, and the corresponding skew Young diagram 
is shown in Fig.\ \ref{fig.SkewYoungDiagram}.
The excitation content 
reads $\bar{\rm s}_{\rm L} + ({\rm s}_{\rm L}^{\uparrow},{\rm h}_{\rm L})
 + \bar{\rm h}^{\ast} + \bar{\rm s}_{\rm R}^{\ast}$.
Then the change of the $z$-component of spin becomes $-1$ and 
there is no change of charge, 
which is consistent with the changes between the initial state and 
the final one.

As the number of sites increases with electron density and spin-polarization
density kept constant, 
the effect of the edges in the holon condensate and 
that in the up-spinon condensate 
should be negligible.
This means that both the half-antiholon and the half-antispinon 
have zero energy.
In fact, for $N(q,\omega)$ at arbitrary filling and zero field, 
the energy levels with contents $\bar{\rm h}^{\ast} + {\rm h}_{\rm R}$ 
are almost along the holon dispersion for $0 \le q \le \bar{n} \pi/2$ 
in the compact support.~\cite{SK}
For $S^{zz}(q,\omega)$ at $\bar{n}=1$ and arbitrary magnetization, 
the energy levels with contents ${\rm s} + \bar{\rm s}^{\ast}_{\rm R}$ 
are almost along the spinon dispersion for $0 \le q \le (1 - \bar{m}) \pi$
in the support by Talstra and Haldane.~\cite{TH}
Thus half-antiholons and half-antispinons should not survive 
in the thermodynamic limit.
In this way, the excitation contents in the thermodynamic limit are 
identified as shown in Table \ref{econtent}.
The $z$-component of spin ($1/2$ or $-1/2$) of spinons is assigned 
so that the spin conservation is satisfied.

When we draw the support using the excitation contents 
(quasi-particles are identified as free particles), 
the estimated region is much larger than the region 
where the poles with finite intensity are present.
This implies the importance of selection rules.
Ha and Haldane~\cite{HH} found the empirical rule in zero magnetic field: 
for a given spinon-holon pair (s$_{\rm R}$, h$_{\rm R}$), 
$|v_{{\rm s}_{\rm R}}| \ge |v_{{\rm h}_{\rm R}}|$, 
and the same to the left-going pair.
Here $v$ is given by $\partial \epsilon / \partial q$.
We find that this rule applies also to the case of nonzero magnetic field.
Then the estimated region is 
found to be the same as 
the region expected from the recursion method.
\par

\subsection{Characteristic features of the supports}

Figures \ref{fig.HSs-+qwmag}-\ref{fig.IStjnqwmag} show the regions 
of compact support 
for $S^{-+}(q,\omega)$, $S^{zz}(q,\omega)$, $S^{+-}(q,\omega)$
and $N(q,\omega)$ with finite magnetization.
The solid lines mean either the spectrum of a quasi-particle 
or that of a pair moving with the same velocity.
The presence of other quasi-particles excited at an edge of each band 
causes several lines to appear with particular shifts in momentum and energy.
Outermost dispersion lines correspond to the boundary of the compact support.
Namely, there is no intensity outside the outermost lines.
\par

The number of zero modes is increased by the external magnetic field.
Namely 
for $S^{-+}(q,\omega)$ and $S^{+-}(q,\omega)$, $\omega=0$ is allowed 
at $q = \bar{m} \pi$, $\bar{n} \pi$, $(2 - \bar{n}) \pi$ and 
$(2 - \bar{m}) \pi$.
For $S^{zz}(q,\omega)$, on the other hand, $\omega=0$ is allowed at 
$q=0$ [$2 \pi$], $2 \bar{m} \pi$, 
$(\bar{n} - \bar{m}) \pi$ [$= v_{\rm s}^0 - v_{\rm c}^0$], 
$(2 - \bar{n} - \bar{m}) \pi$ [$= v_{\rm s}^0 + v_{\rm c}^0$], 
$(\bar{n} + \bar{m}) \pi$, 
$(2 - \bar{n} + \bar{m}) \pi$ and $(2 - 2 \bar{m}) \pi$.
For $N(q,\omega)$, $\omega=0$ is allowed at 
$q=0$ [$2 \pi$], $(2 - 2 \bar{n}) \pi$, 
$(\bar{n} - \bar{m}) \pi$, 
$(2 - \bar{n} - \bar{m}) \pi$, 
$(\bar{n} + \bar{m}) \pi$, $(2 - \bar{n} + \bar{m}) \pi$ 
and $2 \bar{n} \pi$.
\par

For $S^{-+}(q,\omega)$, the excitation contents is the same as 
those for $S(q,\omega)$ in zero magnetic field~\cite{HH} 
(see Table \ref{econtent}).
Note that no antispinons contribute to $S^{-+}(q,\omega)$.
At arbitrary filling, the support turns out to be essentially a 
squeezed version of $S(q,\omega)$ in zero field.

For $S^{zz}(q,\omega)$ and $N(q,\omega)$, the region of 
(s$_{\rm L}$, h$_{\rm L}$) + $\bar{\rm h}$ + (h$_{\rm R}$, s$_{\rm R}$) 
shrinks and splits into three directions by a magnetic field: 
the two of them is the parallel shift by $\pm \bar{m} \pi$ 
along the momentum axis (horizontal axis), 
and the remaining one is the parallel shift by $+h$ 
along the energy axis (vertical axis).
This is due to the presence of the antispinon.

Finally we note that the support of $S^{+-}(q,\omega)$ over $h$ agrees with 
that of $S^{zz}(q,\omega)$ for the same filling and magnetization.

%--------------------------------------------------------------------------

\section{Spectral Weights}

In this section we focus on spectral weights in the compact supports.

\subsection{$S^{-+}(q,\omega)$}

First of all, we discuss the results of $S^{-+}(q,\omega)$ at $\bar{n}=1$.
Figures \ref{fig.HSs-+qwmag}(a) and \ref{fig.HSs-+qwmag}(b) show 
the results in the 16-site chain for
$\bar{m} = 0.25$ (10 up-spins and 6 down-spins) and 
$\bar{m} = 0.5$ (12 up-spins and 4 down-spins), respectively.
With use of $h = \bar{m} (2 - \bar{m})\pi^2 /2$, 
the values of the magnetic field are taken as $h=2.15898$ for $\bar{m}=0.25$ 
and $h=3.70110$ for $\bar{m}=0.5$.
We find that for finite systems the momentum $q$, the excitation energy 
$\omega_{\nu}$ and the form factor $M_{\nu}^q$ 
have strong correspondence with those for $S(q,\omega)$ 
in zero field.
%~\cite{YSAK1,YSAK2}
Namely, for $0 \le c_2 < c_1 \le M$,
with $c_1$ and $c_2 $ being integers, 
it is known that~\cite{YSAK1,YSAK2}
\bea
  q            &=& \bar{m} \pi + \frac{2 \pi}{N} \left( c_1 + c_2 \right),
\label{S-+qwmomentum} \\
  \omega_{\nu} &=& \frac{t}{2} \left( \frac{2 \pi}{N} \right)^2 \bigl[ 
2 M \left( c_1 + c_2 \right) - 2 \left( c_1^2 + c_2^2
\right) \nonumber \\
               & & + c_1 - c_2 \bigr],\label{S-+qwenergy} \\
  M_{\nu}^q    &=& \left\{
   \begin{array}{ll}
    \dfrac{2 c_1 - 2 c_2 - 1}{2 c_1 - 1} \cdot
    \dfrac{1}{2 M - 2 c_2 - 1} \\
    \times \displaystyle{\prod_{i = c_2 + 1}^{c_1 - 1}} \left
    ( \dfrac{2 i}{2 i - 1} \cdot \dfrac{2 M - 2 i}{2 M - 2 i -
    1} \right), \\
    \qquad \qquad \qquad \qquad \mbox{for $c_2 < c_1 - 1$},\\
    \dfrac{1}{\left( 2 c_1 - 1 \right) \left( 2 M - 2 c_1 + 1
    \right)}, \\
    \qquad \qquad \qquad \qquad \mbox{for $c_2 = c_1 - 1$}.
   \end{array}\right.\label{S-+qwformfactor}
%  M_{\nu}^q &=& \frac{1}{2} \left( c_1 - c_2 - \frac{1}{2} \right) 
%\nonumber \\
%            &\times& \prod_{j=1}^2 
%\frac{\Gamma[c_j + (j-1)/2] \Gamma[M - c_j + (2-j)/2]}
%{\Gamma[c_j + j/2] \Gamma[M - c_j + (3-j)/2]}.\qquad \label{S-+qwformfactor}
\eea
Here $M$ denotes the number of down-spins in the initial state.
The number of poles with finite intensity is $M ( M + 1 )/2$.
We have checked the validity of 
eqs.\ (\ref{S-+qwmomentum})-(\ref{S-+qwformfactor}) 
by comparison with numerical results up to $N=16$.
\par

Taking the thermodynamic limit, we obtain the explicit formula
\bea
  & & S^{-+}(q,\omega) \nonumber \\
  & & = \frac{1}{2} \cdot \frac{\Theta \left
    ( \epsilon_{\rm U} (q) - \omega \right) \Theta \left
    ( \omega - \epsilon_{\rm L-} (q) \right) \Theta \left( \omega -
    \epsilon_{\rm L+} (q) \right)}{\sqrt{ \left( \omega -
      \epsilon_{\rm L-} (q) \right) \left( \omega - \epsilon_{\rm L+}
      (q) \right)}},\qquad \label{S-+qomega}
\eea
where $\Theta(\omega)$ is the step function, and
\bea
  \epsilon_{\rm L-}(q) &=& t (q - \bar{m} \pi) (\pi - q),\\
  \epsilon_{\rm L+}(q) &=& t (q - 2 \pi + \bar{m} \pi) (\pi - q),\\
  \epsilon_{\rm U}(q)  &=& \frac{t}{2} (q - 2 \pi + \bar{m} \pi) (\bar{m} 
  \pi - q).
\eea
In the zero-field limit ($\bar{m} = 0$), this
expression reduces to the result obtained by Haldane and Zirnbauer.~\cite{HZ}
\par

Next we consider the case of $\bar{n} < 1$.
Figures \ref{fig.IStjs-+qwmag}(a) and \ref{fig.IStjs-+qwmag}(b) show 
the results of $S^{-+}(q,\omega)$ 
in the 16-site chain with $\bar{n}=0.875$ (2 holes) 
for $\bar{m}=0.25$ (9 up-spins and 5 down-spins) and 
$\bar{m}=0.5$ (11 up-spins and 3 down-spins), respectively.
Like the zero-field case ($\bar{m} = 0$),~\cite{SK} 
at fixed $\bar{m}( > 0)$  
energy levels and intensities in the two-spinon region 
(i.e., the hatched area in Fig.\ \ref{fig.IStjs-+qwmag}) agree 
with those for $\bar{n} = 1$ within the numerical accuracy.
We have checked this fact for $N \le 16$ with various $\bar{n}$ and
$\bar{m}$.
From this fact, it is conjectured that analytical expression of
$S^{-+}(q,\omega)$  in the 2s$_{\rm R}$ and 2s$_{\rm L}$ is identical
with eq.\ (\ref{S-+qomega}).
This feature is an indication of the strong spin-charge separation,
which should be due to supersymmetric Yangian symmetry of 
the present model.
For the nearest-neighbor supersymmetric {\it t-J} model, such strong 
separation does not occur.
This statement holds also for $S^{zz}(q,\omega)$, $S^{+-}(q,\omega)$ 
and $N(q,\omega)$ to be discussed later.
\par

For finite systems, the momentum, the excitation energy and
the form factor are expressed as follows:
\bea
  q              &=& \bar{m} \pi + \frac{2 \pi}{N} c,\\
  \omega_{\nu}   &=& \frac{t}{2} \left( \frac{2 \pi}{N} \right)^2
  f_{\nu} (Q,M),\\
  M_{\nu}^q      &=& g_{\nu} (Q,M).
\eea
Here $M$ and $Q$ denote the numbers of down-spins and holes in the 
initial state, respectively, and $c$ is an integer which satisfies  
$0 < c < Q + 2M$.
We note that $f_{\nu}(Q,M)$ and $g_{\nu}(Q,M)$ are
 independent of the system size $N$.
This implies that $S^{-+}(q,\omega)$ for the present model is 
equivalent to the hole propagator, where one boson is removed, 
for the SU(1,1) Calogero-Sutherland model 
with $Q$ fermions and $M$ bosons.~\cite{Polychronakos,KK94}
\par

\subsection{$S^{zz}(q,\omega)$}

Figures \ref{fig.HSszzqwmag}(a) and \ref{fig.HSszzqwmag}(b) show 
the results of $S^{zz}(q,\omega)$ for the 16-site 
Haldane-Shastry model with magnetization $\bar{m} = 0.25$ and $0.5$, 
respectively.
The intensity at $(q,\omega)=(0,0)$ is given by $N \bar{m}^2/4$.
It is found that dominant intensity lies in the region surrounded 
by four lines: A, B, C and D in Fig.\ \ref{fig.HSszzqwmag}.
Analytical expression of $S^{zz}(q,\omega)$ 
in the region $q \le \bar{m} \pi$ has been computed with use of 
strong coupling limit of the spin Calogero-Sutherland model.~\cite{ASYK}
\par

In the 16-site chain with $\bar{n}=0.875$, the results of $S^{zz}(q,\omega)$ 
for $\bar{m}=0.25$ and $0.5$ are shown in Figs.\ \ref{fig.IStjszzqwmag}(a) 
and \ref{fig.IStjszzqwmag}(b), respectively. 
The compact support reveals the region where only quasi-particles 
with spin degrees of freedom (i.e., spinon and antispinons) contribute.
In this region, energy levels and intensities agree with those for 
$\bar{n}=1$ with the same $\bar{m}$.
This is an indication of the strong spin-charge separation in 
the longitudinal component of spin dynamics.
\par

\subsection{$S^{+-}(q,\omega)$}

In Figs.\ \ref{fig.HSs+-qwmag}(a) and \ref{fig.HSs+-qwmag}(b) 
we show the results of $S^{+-}(q,\omega)$ 
in the 16-site chain at $\bar{n}=1$ for
$\bar{m} = 0.25$ and $0.5$, respectively.
At $q=0$ there is a single pole; its energy level is $\omega = h$ and
its intensity $| \langle \nu | S^-_{q=0} | 0 \rangle |^2 = \bar{m}$.
The lower edge over $0 \le q \le \bar{m} \pi$ is not only the boundary 
of the continuum but also the dispersion line of an antispinon (i.e.,
magnon) excited alone.~\cite{TH}

At a fixed $\bar{n}=0.875$ we show the results for the 16-site chain
with $\bar{m}=0.25$ and $0.5$ in Figs.\ \ref{fig.IStjs+-qwmag}(a) 
and \ref{fig.IStjs+-qwmag}(b), respectively.
Let us compare the present case with the half-filled case in the same
magnetization.
Although the structure of intensity is partially broken down 
by existence of holes, the $(q,\omega)$-region lies 
where the structure remains intact.
Namely, energy levels and intensities in the region agree with those 
for $\bar{n}=1$ and the same magnetization.
This region corresponds to a case where 
holons and antiholons are never excited, in terms of the support in
the thermodynamic limit.
This fact indicates the strong spin-charge separation in
$S^{+-}(q,\omega)$.

%--------------------------------------------------------------------------

\subsection{$N(q,\omega)$}

We now turn to the charge dynamics.
In ref.\ 18 we have discussed the features of $N(q,\omega)$ 
in zero magnetic field.
Figures \ref{fig.IStjnqwmag}(a) and \ref{fig.IStjnqwmag}(b) show our results for electron 
density $\bar{n} = 0.875$ (16 sites and 2 holes) with 
two values of magnetization $\bar{m}$.
The following remarkable feature is found: 
in the pure $\bar{\rm h}$ + 2h$_{\rm R}$ region, 
the poles and intensities of $N(q,\omega)$ are independent of 
$\bar{m}$ within the numerical accuracy.
For $q \le (\bar{n}-\bar{m})\pi/2$, in particular, this statement 
has been proved analytically.~\cite{AYSK} 
The independence is a manifestation of the strong spin-charge separation 
in charge dynamics.

%%%%%%%%%%%%%%%%%%%%%%%%%%%%%%%%%%%%%%%%%%%%%%%%%%%%%%%%%%%%%%%%%%%%%%%%%%%
%%%%

\section{Static Structure Factors}

Integration of the dynamical structure factor over $\omega$ yields
the static structure factor.
Figures \ref{fig.IStjszzqmag}-\ref{fig.IStjnqmag} show numerical 
results of $S^{zz}(q)$, $S^{xx}(q)$ and 
$N(q)$ for the inverse-square type.

Particularly for $S^{xx}(q)$, the analytic expression can be obtained.
At $\bar{n}=1$, by integrating eq.\ (\ref{S-+qomega}) over $\omega$, 
we obtain
\be
  S^{-+}(q) = \left\{
   \begin{array}{ll}
    0, &\mbox{for $0 \le q \le \bar{m} \pi$},\\
    - \dfrac{1}{2} \log \dfrac{1 - q/\pi}{1 - \bar{m}}, &\mbox{for
      $\bar{m} \pi \le q \le \pi$}.
   \end{array}\right.\label{S-+qhf}
\ee
The Fourier transformation leads to
\be
  \langle S_0^- S_r^+ \rangle = \frac{(-1)^r}{2 \pi r} {\rm Si} \left( (1 - 
  \bar{m}) \pi r \right),
\ee
where $r$ is an integer which satisfies $r \ge 1$, 
and ${\rm Si} (x)$ is the sine integral.
\par

For $\bar{n}<1$, numerical calculations show that 
$S^{-+}(q)$ are equivalent to that for $\bar{n}=1$ 
if $0 \le q \le \bar{n} \pi$; 
$S^{-+}(q)$ for $\bar{n} \pi \le q \le \pi$ are equal to 
the value for $q = \bar{n} \pi$.
On the basis of these facts, we conjecture the following expression:
\be
  S^{-+}(q) = \left\{
   \begin{array}{ll}
    0, &\mbox{for $0 \le q \le \bar{m} \pi$},\\
    - \dfrac{1}{2} \log \dfrac{1 - q/\pi}{1 - \bar{m}}, &\mbox{for
      $\bar{m} \pi \le q \le \bar{n} \pi$},\\
    - \dfrac{1}{2} \log \dfrac{1 - \bar{n}}{1 - \bar{m}}, &\mbox{for
      $\bar{n} \pi \le q \le \pi$}.\label{S-+qlhf}
   \end{array}\right.
\ee
The Fourier transformation leads to
\be
  \langle S_0^- S_r^+ \rangle = \frac{(-1)^r}{2 \pi r} \left[ {\rm Si} 
  \left( (1 - \bar{m}) \pi r \right) - {\rm Si} \left( (1 - \bar{n})
    \pi r \right) \right].
\ee
When $\bar{m}=0$, this result reduces to the previous result.~\cite{GV,YO}

The relation between $S^{-+}(q)$ and $S^{+-}(q)$ is as follows:
\be
  S^{+-}(q) = \bar{m} + S^{-+}(q),
\ee
irrespective of the electron density.
Therefore, using the relation $S^{xx}(q) = [ S^{-+}(q) + S^{+-}(q) ]/4$, 
we get analytic expression of $S^{xx}(q)$:
\be
  S^{xx}(q) = \frac{1}{4} \bar{m} + \frac{1}{2} S^{-+}(q),
\ee
where $S^{-+}(q)$ is given by eq.\ (\ref{S-+qhf}) or eq.\ (\ref{S-+qlhf}).

The results of static structure factors for the inverse-square type 
are summarized as follows:

(i) At fixed $\bar{m}$, $S^{zz}(q)$ is independent of $\bar{n}$
for $0 \le q \le (\bar{n} - \bar{m}) \pi$.

(ii) At fixed $\bar{m}$, $S^{xx}(q)$ is independent of $\bar{n}$
for $0 \le q \le \bar{n} \pi$.

(iii) At fixed $\bar{n}$, $N(q)$ is independent of $\bar{m}$
for $0 \le q \le (\bar{n} - \bar{m}) \pi$.

\noindent These features are regarded as a manifestation of the 
strong spin-charge separation in static structure factors.

What about the case of the nearest-neighbor interaction?
Figures \ref{fig.NNtjszzqmag}, \ref{fig.NNtjsxxqmag} and \ref{fig.NNtjnqmag} 
show the results of $S^{zz}(q)$, $S^{xx}(q)$
and $N(q)$, respectively, for the nearest-neighbor type 
with two representative values of $J/t$. 
We look at the difference of $S^{\alpha \alpha}(q)$ [$\alpha=z,x$] 
between the two values of filling ($\bar{n}=1$ and $0.875$), 
and that of $N(q)$ between the two 
values of magnetization ($\bar{m}=0$ and $0.25$). 
In any case, within the special momentum range described in (i)-(iii), 
the difference is less than 6\% for $S^{zz}(q)$, 5\% for $S^{xx}(q)$ 
($q = \bar{n}\pi$ not included), and 1\% for $N(q)$.
This indicates that the 1D {\it t-J} model with realistic parameters 
shares the properties of static structure factors in the $1/r^2$-type 
supersymmetric {\it t-J} model.
The independence of static spin (charge) structure factors with 
varying filling (magnetization) spreads over considerably wide 
range of momentum, if the system is near half-filling and under a 
weak magnetic field.
This behavior might be detected experimentally
in a quasi-1D conductor.

%%%%%%%%%%%%%%%%%%%%%%%%%%%%%%%%%%%%%%%%%%%%%%%%%%%%%%%%%%%%%%%%%%%%%%%%%%%

\section{Summary and Discussion} \label{Summary}

We have investigated the dynamical and static properties of the 
1D supersymmetric {\it t-J} model with $1/r^2$ interaction
in a magnetic field.
We have determined the $(q,\omega)$-region of nonvanishing
spectral weight for $S^{\alpha \alpha}(q,\omega)$ [$\alpha = x,y,z$] 
and $N(q,\omega)$ in nonzero magnetic field.
We have found the following features about the spectral weights 
themselves:

(1) At fixed $\bar{m}$, $S^{\alpha \alpha}(q,\omega)$ is independent of
$\bar{n}$ in the region where only quasi-particles with spin degrees
of freedom (spinons and antispinons) contribute.

(2) At fixed $\bar{n}$, $N(q,\omega)$ is independent of $\bar{m}$ in
the region where only quasi-particles with charge (holons and
antiholons) contribute.

These features 
constitute further manifestation of
the strong spin-charge separation
in the $1/r^2$-type supersymmetric {\it t-J} model. 

When a system deviates from the $1/r^2$-type model, 
elementary excitations in a magnetic field do not correspond to 
spinons and holons but to mixtures of them.~\cite{Bares,Carmelo,Pham}
This is related to the fact that the system loses the $Z_2$ symmetry between 
the statistical properties of up- and down-spin electrons in the presence of 
finite polarization.~\cite{Pham}
In fact we have checked for the polarized nearest-neighbor {\it t-J} model 
that the long-wavelength limit
of the spin and charge excitations have a common velocity. 
Thus the spin-charge separation is not realized any longer. 
On the other hand, it has been shown that spinons and holons together with 
their antiparticles span the complete set in the Fock space of 
hard-core lattice fermions with any 
filling and polarization~\cite{KK1,KK2}.
The completeness is independent of whether or not spinons and holons form 
eigenstates of the model. 
Thus the absence of spin-charge separation in the polarized 
nearest-neighbor {\it t-J} model means that there are residual interactions 
between spinons and holons.

In contrast, the $1/r^2$-type model maintains the $Z_2$ invariance 
even with finite polarization.
This can be understood from the statistical matrix given in ref.\ 16. 
The parallel dispersion relations of up and down spinons as shown in 
Fig.\ \ref{fig.spectra}(a) are also consistent with the $Z_2$ invariance. 
As we have shown in this paper, spinons and holons remain elementary 
excitations, and are essential to understanding dynamics 
for any polarization in the $1/r^2$-type supersymmetric {\it t-J} model. 
This property entitles the model to be called a ``canonical'' or 
a kind of ``free'' Hamiltonian, 
which is nothing but another representation of the higher symmetry 
in the model.

\section*{Acknowledgments}

We thank M. Imada for his support and helpful comments on this work.
We also acknowledge fruitful discussions with Y. Kato, 
T. Yamamoto and M. Arikawa.
The numerical calculations were performed partly at the Supercomputer 
Center of the Institute for Solid State Physics, University of Tokyo.

\appendix
\section{Derivation of Dispersion Relations of Quasi-Particles}

In this appendix, we derive dispersion relations of four types 
of quasi-particles (i.e., spinons, antispinons, holons and antiholons)
for the $1/r^2$-type supersymmetric {\it t-J} model in a magnetic
field.
The derivation is based on the asymptotic Bethe-ansatz equations.

Let us take the completely up-polarized state as the reference state 
and consider $M+Q$ pseudo-particles 
which represent $M$ electrons with spin down and $Q$ holes.
The asymptotic Bethe ansatz leads to the following 
equation:~\cite{Kawakami,WLC}
\be
  \frac{E}{t} = \sum_{\mu = 1}^{M+Q} \epsilon (p_{\mu})
+ \frac{\pi^2}{3} Q \left( 1 - \frac{1}{N^2} \right)
- h \left( \frac{N-Q}{2} - M \right),
\label{EnergyABA}
\ee
where 
$\epsilon(p) = p (p - 2 \pi)/2$ 
for $0 \le p \le 2\pi$.
For later convenience the origin of pseudomomenta is shifted 
by $\pi$.
Then $\bar{\epsilon}(p) \equiv \epsilon (p+\pi)
= (p^2 - \pi^2)/2$ for $|p| \le \pi$.
When we consider the value of the physical momentum, 
we restore the origin of pseudomomenta.
Note that eq.\ (\ref{EnergyABA}) includes the Zeeman term.
The pseudomomenta $p_{\mu}$ are determined by the following equations:
\bea
  2 \pi J_{\mu} &=&  p_{\mu} N + \pi \sum_{i=1}^Q {\rm sgn} (p_{\mu} - q_i) 
\nonumber \\
            & & - \pi \sum_{\nu = 1}^{M+Q} {\rm sgn} (p_{\mu} -
            p_{\nu})
            \equiv  z(p_\mu) N,
            \label{momentumABA1} \\
  2 \pi I_i &=& \pi \sum_{\mu = 1}^{M+Q} {\rm sgn} (q_i - p_{\mu}) 
  \equiv  w(q_i) N,\label{momentumABA2}
\eea
where $J_{\mu} \in [-(N-M-1)/2, (N-M-1)/2]$ for $\mu=1, 2, \cdots, M+Q$,
and $I_i \in [-(M+Q)/2, (M+Q)/2]$ for $i=1, 2, \cdots, Q$.
The quantum numbers $J_\mu$ and $I_i$ are integers or half-integers, 
and are arranged in the ascending order.

Before considering the dispersion relations, we must derive 
the ground-state distribution functions of pseudomomenta.
For the ground state, both $J_{\mu}$ and $I_i$ are
distributed densely and symmetrically with respect to zero.
Namely, in the subspace with $(N,Q,M)$, 
$J_\mu = - (M+Q-1)/2 + \mu - 1$ for $\mu=1,2,\cdots,M+Q$, 
and $I_i = - (Q-1)/2 + i - 1$ for $i=1,2,\cdots,Q$. 
In the large-$N$ limit, using $2 \pi \sigma(p) = d z(p)/ d p$ and 
$2 \pi \rho(q) = d w(q)/ d q$,
we obtain
\bea
  \sigma_0 (p) &=& \Theta ( B_0 - |p| ) \Biggl[ \frac{1}{2 \pi} + 
\int_{-D_0}^{D_0} d q \rho_0 (q) \delta(p-q) \nonumber \\
               & & - \int_{-B_0}^{B_0} d p' \sigma_0 (p') \delta(p-p') \Biggr],
\label{distri1} \\
  \rho_0 (q)   &=& \Theta ( D_0 - |q| ) 
\int_{-B_0}^{B_0} d p \sigma_0 (p) \delta(q-p),
\label{distri2}
\eea
where $\Theta(p)$ is the step function.
Here we have attached the subscript $0$ to quantities corresponding to the ground state.
The cut-off parameters $B_0$ and $D_0$ ($B_0 \ge D_0$) are determined by
\bea
  & & \frac{2 - \bar{n} - \bar{m}}{2} \equiv \frac{M+Q}{N} 
= \int_{-B_0}^{B_0} \sigma_0 (p) d p,\label{cutoff1} \\
  & & 1 - \bar{n} \equiv \frac{Q}{N} = \int_{-D_0}^{D_0} \rho_0 (q) d q.
\label{cutoff2}
\eea
Equations (\ref{distri1})-(\ref{cutoff2}) lead to the ground-state 
distributions $\sigma_0 (p)$ and $\rho_0 (q)$:~\cite{Kawakami}
\bea
  \sigma_0 (p) &=& \left\{
   \begin{array}{ll}
    1/(2 \pi), &\mbox{for $|p| \le (1 - \bar{n}) \pi$},\\
    1/(4 \pi), &\mbox{for $(1 - \bar{n}) \pi \le |p| \le (1 - \bar{m}) \pi$},
\qquad \\
    0, &\mbox{otherwise},
   \end{array}\right.\\
  \rho_0 (q)   &=& \left\{
   \begin{array}{ll}
    1/(2 \pi), &\mbox{for $|q| \le (1 - \bar{n}) \pi$},\\
    0, &\mbox{otherwise}.
   \end{array}\right.
\eea
We then obtain the ground-state energy per site
\bea
  \frac{\epsilon_0}{t} &=& \int_{- \pi}^{\pi} d p \sigma_0 (p) 
\bar{\epsilon}(p) 
+ \frac{\pi^2}{3} (1 - \bar{n}) - \frac{1}{2} \bar{m} h \nonumber \\
             &=& - \frac{\pi^2}{12} \left( \bar{n}^3 - 3 \bar{n}^2
+ 4 \bar{n} + \bar{m}^3 - 3 \bar{m}^2 \right) - \frac{1}{2} \bar{m} h,\qquad
 \quad
\eea
in the thermodynamic limit.~\cite{Kawakami,YK}
The ground-state momentum is given by
\bea
  q_0 &=& N \int_{- \pi}^{\pi} d p \sigma_0 (p) (p + \pi) \nonumber \\
      &=& \frac{\pi N}{2} (2 - \bar{n} - \bar{m}).
\eea

Now that we have obtained information on the ground state, 
the next task is to derive the dispersion relations of quasi-particles.
We take the following steps:

(i) We first give the configurations of $J_\mu$ and $I_i$ so that 
the quasi-particle can be created in the pseudomomentum space.

(ii) We find distribution functions $\sigma(p)$ and $\rho(q)$ under 
the configurations given in (i).

(iii) In general, the distribution functions determining the excitation 
energy can be written in the form
\bea
  \sigma(p) &=& \sigma_0 (p) + (1/N) \sigma_1 (p),\\
  \rho(q)   &=& \rho_0 (q) + (1/N) \rho_1 (q).
\eea
Thereby we get the explicit expression of $\sigma_1 (p)$.

(iv) We obtain the excitation energy and the momentum transfer by using 
$\sigma_1 (p)$, and then derive the dispersion relation.

\subsection{Spinons}

In a magnetic field, the dispersion relation of up spinons should be 
different from that of down spinons due to the Zeeman splitting.
First, in order to obtain the up-spinon dispersion, we consider 
the spin excitation in the subspace with $(N,Q+1,M-1)$, 
where a down spin is removed.
We take the numbers $I_i$ ($i=1,2,\cdots,Q+1$) as 
\be
  I_i = - (Q-1)/2 + i - 1,
\ee
which means that the charge excitation (holon excitation) takes 
the minimum energy
that vanishes in the thermodynamic limit.
On the other hand we choose the configuration of $J_\mu$ 
($\mu = 1,2,\cdots,M+Q)$ 
with $\mu_0$ being the location of spinon excitation 
as
\be
  J_{\mu + 1} - J_\mu = 1 + \delta_{\mu,\mu_0},
\ee
where $J_\mu \in [-(M+Q)/2,(M+Q)/2]$.
The position of a ``hole'' in the $J_\mu$-configuration is denoted by 
$\bar{J}$.
When $-(M+Q)/2 \ll \bar{J} \ll I_1$ or $I_{Q+1} \ll \bar{J} \ll (M+Q)/2$, 
a spinon is well-defined in the pseudomomentum space. 
The distribution functions $\sigma(p)$ and $\rho(q)$ must 
satisfy the following equations:
\bea
  \sigma(p) &=& \Theta ( B - |p| ) \Biggl[ \frac{1}{2 \pi} + 
\int_{-D}^{D} d q \rho(q) \delta(p-q) \nonumber \\
            & & - \int_{-B}^B d p' \sigma (p') \delta(p-p') \nonumber \\
            & & - \frac{1}{N} \delta (p - p_{\rm h}) 
\Theta (|p| - D) \Biggr],\label{distri1upspinon} \\
  \rho(q)   &=& \Theta ( D - |q| ) \int_{-B}^{B} d p \sigma (p) \delta(q-p),
\label{distri2upspinon}
\eea
where $p_{\rm h}$ is the value of $p$ for $\mu=\mu_0$.
Note that these equations fulfill the normalization conditions:
\bea
  \int_{-B}^B \sigma(p) d p &=& \frac{M+Q}{N},\\
  \int_{-D}^D \rho(q) d q   &=& \frac{Q+1}{N}.
\eea

By introducing $\sigma_1 (p)$ by $\sigma(p) = \sigma_0 (p) 
+ (1/N) \sigma_1 (p)$, we find 
\bea
  \sigma_1 (p) &=& - \frac{1}{2} \delta (p - p_{\rm h})
\Theta (|p| - (1-\bar{n})\pi) \nonumber \\
               & & \times \Theta ((1-\bar{m})\pi - |p|).
\label{sigma1upspinon}
\eea
We have used $B \cong B_0$ and $D \cong D_0$ in the large-$N$ limit.
Then the excitation energy from the chemical potential is given by 
\bea
  \epsilon &=& \int_{-\pi}^{\pi} d p \sigma_1 (p) \bar{\epsilon} (p) 
- \frac{h}{2} \nonumber \\
           &=& - \frac{1}{4} (p_{\rm h}^2 - \pi^2) - \frac{h}{2},
\label{energyupspinon}
\eea
where $(1-\bar{n})\pi \le |p_{\rm h}| \le (1-\bar{m})\pi$.
For the associated momentum transfer 
$-k$, we obtain 
\bea
  -k &=& \int_{-\pi}^{\pi} d p \sigma_1 (p) (p + \pi) \nonumber \\
     &=& - \frac{1}{2} (p_{\rm h} + \pi).
\eea
Therefore we obtain eq.\ (\ref{upspinondispersion}).
The constraint on $p_{\rm h}$ results in the 
following 
range of allowed momentum of 
up spinons:
$\bar{m}\pi/2 \le k \le \bar{n}\pi/2$ 
and $(1-\bar{n}/2)\pi \le k \le (1-\bar{m}/2)\pi$.

Next, in order to obtain the down-spinon dispersion, we consider 
the spin excitation in the subspace with $(N,Q+1,M)$, 
where an up spin is removed.
The numbers $I_i$ ($i=1,2,\cdots,Q+1$) are chosen as
\be
  I_i = -Q/2 + i - 1,
\ee
and the configuration of $J_\mu$ ($\mu = 1,2,\cdots,M+Q+1$) are
\be
  J_{\mu + 1} - J_\mu = 1 + \delta_{\mu,\mu_0},
\ee
where $J_\mu \in [-(M+Q+1)/2,(M+Q+1)/2]$.
In a manner similar to that 
for the up spinon, we get the same expression of $\sigma_1 (p)$ 
as given by eq.\ (\ref{sigma1upspinon}).
The second term of eq.\ (\ref{energyupspinon}) is replaced by $+h/2$ 
because the subspace we consider is different from that for the up spinon.

\subsection{Antispinons}

In order to obtain the 
spectrum of antispinons, we consider the excitation 
in the subspace with $(N,Q,M+1)$,
where an up-spin flips into a down-spin.
The numbers $I_i$ ($i=1,2,\cdots,Q$) and $J_\mu$ ($\mu = 1,2,\cdots,M+Q+1$) 
are chosen as follows:
\bea
  & & I_i = -Q/2 + i - 1, \\
  & & J_\mu = -(M+Q)/2 + \mu,\nonumber \\
  & & \hspace{2cm} {\rm for} \ \ \mu=1,2,\cdots,M+Q,\\
  & & J_{M+Q+1} - J_{M+Q} \gg 1,
\eea
or
\bea
  & & I_i = -Q/2 + i, \\
  & & J_\mu = -(M+Q)/2 + \mu - 2,\nonumber \\
  & & \hspace{2cm} {\rm for} \ \ \mu=2,3,\cdots,M+Q+1,\\
  & & J_2 - J_1 \gg 1.
\eea
In the former (latter) case, $p_{\rm p}$ is defined as the value of $p$ for 
$J_\mu = J_{M+Q+1}$ ($J_1$).
The distribution function $\sigma(p)$ can be written in the form
\bea
  \sigma(p) &=& \sigma_0 (p) \Theta ((1-\bar{m})\pi - |p|) \nonumber \\
            & & + \frac{1}{N} \delta (p - p_{\rm p}) 
\Theta (|p| - (1-\bar{m})\pi) \nonumber \\
            & & \times \Theta (\pi - |p|),
\eea
Note that this equation fulfills the normalization condition: 
$\int \sigma(p) d p = (M+Q+1)/N$.
We can identify $\sigma_1 (p)$ as follows:
\bea
  \sigma_1 (p) &=& \delta (p - p_{\rm p})
\Theta (|p| - (1-\bar{m})\pi) \nonumber \\
               & & \times \Theta (\pi - |p|).
\eea
Then the excitation energy is given by
\be
  \epsilon = \frac{1}{2} (p_{\rm p}^2 - \pi^2) + h,
\ee
and the associated momentum transfer is $k = p_{\rm p} + \pi$.
Therefore we obtain the antispinon dispersion given by
eq.\ (\ref{downantispinondispersion}).

\subsection{Holons}

We consider the charge excitation in the subspace with $(N,Q+1,M-1)$, 
where a down-spin is removed from the ground state.
The numbers $J_\mu$ ($\mu=1,2,\cdots,M+Q$) 
are chosen as follows:
\be
  J_\mu = -(M+Q)/2 + \mu - 1, 
\ee
which means that the spin excitation (up-spinon excitation) takes 
the minimum energy that 
vanishes in the thermodynamic limit.
On the other hand we choose $I_i$ ($i=1,2,\cdots,Q+1$) either as
\bea
  & & I_i = -(Q-1)/2 + i - 1, \quad {\rm for} \ \ i=1,2,\cdots,Q,\qquad \\
  & & I_{Q+1} - I_Q \gg 1,
\eea
or as
\bea
  & & I_i = -(Q-1)/2 + i - 2, \nonumber \\
  & & \hspace{2cm} {\rm for} \ \ i=2,3,\cdots,Q+1,\\
  & & I_2 - I_1 \gg 1.
\eea
In the former case we denote $q_{\rm p} = 2 \pi I_{Q+1}/N$, and 
in the latter case $q_{\rm p} = 2 \pi I_1 /N$.
The distribution function $\rho(q)$ can be written in the form
\bea
  \rho(q) &=& \frac{1}{2 \pi} \Theta ((1-\bar{n})\pi - |q|) \nonumber \\
          & & + \frac{1}{N} \delta (q - q_{\rm p}) 
\Theta (|q| - (1-\bar{n})\pi).
\eea
On the other hand, the distribution function $\sigma(p)$ satisfies the 
following equation:
\bea
  \sigma(p) &=& \Theta ( B - |p| ) \Biggl[ \frac{1}{2 \pi} + 
\int_{-D}^{D} d q \rho(q) \delta(p-q) \nonumber \\
            & & - \int_{-B}^B d p' \sigma (p') \delta(p-p') \Biggr].
\label{distri1holon}
\eea
Note that these equations fulfill the normalization conditions: 
$\int \rho(q) d q = (Q+1)/N$ and $\int \sigma(p) d p = (M+Q)/N$.

By introducing $\sigma_1 (p)$ by $\sigma(p) = \sigma_0 (p) 
+ (1/N) \sigma_1 (p)$, we find 
\bea
  \sigma_1 (p) &=& \frac{1}{2} \delta (p - q_{\rm p})
\Theta (|p| - (1-\bar{n})\pi) \nonumber \\
               & & \times \Theta ((1-\bar{m})\pi - |p|).
\eea
We have used $B \cong B_0$ in the large-$N$ limit.
Then the excitation energy is given by
\be
  \epsilon = \frac{1}{4} (q_{\rm p}^2 - \pi^2) + \frac{\pi^2}{3}.
\ee
The second term comes from $\pi^2 Q (1 - 1/N^2)/3$ 
of eq.\ (\ref{EnergyABA}).
The momentum transfer is given by $k = (q_{\rm p} + \pi)/2$.
The relation between $\epsilon$ and $k$ leads to 
eq.\ (\ref{holondispersion2}).
The momentum $k$ is allowed for $\bar{m}\pi/2 \le k \le \bar{n}\pi/2$ 
and $(1-\bar{n}/2)\pi \le k \le (1-\bar{m}/2)\pi$.
The lowest excitation energy, which 
occurs at $k=\bar{n}\pi/2$ 
or $(1-\bar{n}/2)\pi$, corresponds to the absolute value of the 
chemical potential.~\cite{KY,SK}

\subsection{Antiholons}

In order to obtain the antiholon dispersion, we consider the excitation 
where two electrons are added.
We assume that one of the two electrons has up-spin and the other has 
down-spin.
Then the subspace becomes $(N,Q-2,M+1)$.
The numbers of $J_\mu$ ($\mu=1,2,\cdots,M+Q-1$) are chosen as
\be
  J_\mu = -(M+Q)/2 + \mu,
\ee 
and the configurations of $I_i$ ($i=1,2,\cdots,Q-2$) are
\be
  I_{i+1} - I_i = 1 + \delta_{i,i_0},
\ee
where $I_i \in [-(Q-2)/2,(Q-2)/2]$.
The position of a ``hole'' in the $I_i$-configuration is denoted 
by $\bar{I}$.
When $-(Q-2)/2 \ll \bar{I} \ll (Q-2)/2$, an antiholon is well-defined 
in the pseudomomentum space.
The distribution function $\rho(q)$ can be written in the form
\be
  \rho(q) = \left[ \frac{1}{2 \pi} - 
\frac{2}{N} \delta (q - q_{\rm h}) \right] 
\Theta ((1-\bar{n})\pi - |q|),
\ee
where $q_{\rm h}$ is the value of $q$ for $i = i_0$.
Note that this equation fulfills the normalization condition: 
$\int \rho(q) d q = (Q-2)/N$.

By introducing $\sigma_1 (p)$ by $\sigma(p) = \sigma_0 (p) 
+ (1/N) \sigma_1 (p)$, we find
\be
  \sigma_1 (p) = - \delta (p - q_{\rm h}) 
\Theta ((1-\bar{n})\pi - |p|).
\ee
Then the excitation energy is given by
\be
  \epsilon = - \frac{1}{2} (q_{\rm h}^2 - \pi^2) - \frac{2}{3} \pi^2.
\ee
The second term comes from $\pi^2 Q (1 - 1/N^2)/3$ 
of eq.\ (\ref{EnergyABA}).
The momentum transfer is given by $- k = - (q_{\rm h} + \pi)$.
Therefore we obtain eq.\ (\ref{antiholondispersion2}).

%\bigskip
\newpage

% table

\begin{table}[hbtp]
\caption{List of all the possible excitations from the ground state 
with finite magnetization.}
\label{econtent}
\end{table}

% Caption of figures
\begin{figure}
\caption{Dispersion relations of (a) up and down spinons, (b) antispinons, 
(c) holons and (d) antiholons 
as given by eqs.\ (\ref{upspinondispersion})-(\ref{antiholondispersion2}).
Allowed ranges of momenta and the right and left branches are shown for each 
species.}
\label{fig.spectra} 
\end{figure}

\begin{figure}
\caption{Skew Young diagram corresponding to the motif 
$010|10|110|10|100$.
There are other possibilities to put $1,2$ or $\circ$ than 
the one shown in this figure.}
\label{fig.SkewYoungDiagram}
\end{figure}

\begin{figure}
%\vspace*{13cm}
\caption{$S^{-+}(q,\omega)/4$ of the Haldane-Shastry model with 16 sites 
and (a) $\bar{m} = 0.25$ and (b) $\bar{m} = 0.5$.
The intensity of each pole is proportional to the area of the circle.
The solid lines show the dispersion lines of quasi-particles 
in the thermodynamic limit.
This way of representation applies to all figures to follow.
}
\label{fig.HSs-+qwmag}
\end{figure}

\begin{figure}
%\vspace*{13cm}
\caption{$S^{-+}(q,\omega)/4$ of the $1/r^2$-type supersymmetric 
{\it t-J} model with 16 sites, 2 holes and 
(a) $\bar{m} = 0.25$ and (b) $\bar{m} = 0.5$.
The hatched area indicates the region where only two spinons 
contribute.
}
\label{fig.IStjs-+qwmag}
\end{figure}

\begin{figure}
%\vspace*{13cm}
\caption{$S^{zz}(q,\omega)$ of the Haldane-Shastry model with 16 sites 
and (a) $\bar{m} = 0.25$ and (b) $\bar{m} = 0.5$.
The broken line indicates the magnetic field $h$.
}
\label{fig.HSszzqwmag}
\end{figure}

\begin{figure}
%\vspace*{13cm}
\caption{$S^{zz}(q,\omega)$ of the $1/r^2$-type supersymmetric 
{\it t-J} model with 16 sites, 2 holes and 
(a) $\bar{m} = 0.25$ and (b) $\bar{m} = 0.5$.
The broken line indicates the magnetic field $h$.
The hatched area indicates the region where holons and antiholons do not 
contribute at all.
}
\label{fig.IStjszzqwmag}
\end{figure}

\begin{figure}
%\vspace*{13cm}
\caption{$S^{+-}(q,\omega)/4$ of the Haldane-Shastry model with 16 sites 
and (a) $\bar{m} = 0.25$ and (b) $\bar{m} = 0.5$.
The broken line indicates the magnetic field $h$ and its double $2h$.
}
\label{fig.HSs+-qwmag}
\end{figure}

\begin{figure}
%\vspace*{13cm}
\caption{$S^{+-}(q,\omega)/4$ of the $1/r^2$-type supersymmetric 
{\it t-J} model with 16 sites, 2 holes and 
(a) $\bar{m} = 0.25$ and (b) $\bar{m} = 0.5$.
The broken line indicates the magnetic field $h$ and its double $2h$.
The hatched area indicates the region where holons and antiholons do not 
contribute at all.
}
\label{fig.IStjs+-qwmag}
\end{figure}

\begin{figure}
%\vspace*{13cm}
\caption{$N(q,\omega)$ of the $1/r^2$-type supersymmetric 
{\it t-J} model with 16 sites, 2 holes and 
(a) $\bar{m} = 0.25$ and (b) $\bar{m} = 0.5$.
The broken line indicates the magnetic field $h$.
The hatched area indicates the region where spinons and antispinons do not 
contribute at all.
}
\label{fig.IStjnqwmag}
\end{figure}

\begin{figure}
%\vspace*{13cm}
\caption{$S^{zz}(q)$ of the $1/r^2$-type supersymmetric {\it t-J} model 
with 16 sites, $\bar{m} = 0.25$ and 
two different values of filling.
}
\label{fig.IStjszzqmag}
\end{figure}

\begin{figure}
%\vspace*{13cm}
\caption{$S^{xx}(q)$ of the $1/r^2$-type supersymmetric {\it t-J} model 
with 16 sites, $\bar{m} = 0.25$ and two different values of filling.
}
\label{fig.IStjsxxqmag}
\end{figure}

\begin{figure}
%\vspace*{13cm}
\caption{$N(q)$ of the $1/r^2$-type supersymmetric {\it t-J} model 
with 16 sites, $\bar{n} = 0.875$ and two different values of magnetization.
}
\label{fig.IStjnqmag}
\end{figure}

\begin{figure}
%\vspace*{13cm}
\caption{$S^{zz}(q)$ of the nearest-neighbor-type {\it t-J} model 
with 16 sites, $\bar{m} = 0.25$ and two different values of filling. 
(a) $J/t = 2$; (b) $J/t = 0.5$.
}
\label{fig.NNtjszzqmag}
\end{figure}

\begin{figure}
%\vspace*{13cm}
\caption{$S^{xx}(q)$ of the nearest-neighbor-type {\it t-J} model 
with 16 sites, $\bar{m} = 0.25$ and two different values of filling.
(a) $J/t = 2$; (b) $J/t = 0.5$.
}
\label{fig.NNtjsxxqmag}
\end{figure}

\begin{figure}
%\vspace*{13cm}
\caption{$N(q)$ of the nearest-neighbor-type {\it t-J} model 
with 16 sites, $\bar{n} = 0.875$ and two different values of magnetization.
(a) $J/t = 2$; (b) $J/t = 0.5$.
}
\label{fig.NNtjnqmag}
\end{figure}


\begin{thebibliography}{99}
\bibitem{comment}
In refs.\ 17 and 18, charges of the holon 
and the antiholon have been assigned as $- {\rm e}$ and $+ 2{\rm e}$, 
respectively.
\end{thebibliography}
\end{document}